\documentclass[12pt]{article}

\newcommand{\add}{\addtocounter{eqncnt}{1}}
\newcounter{eqncnt}[section]

\newcommand{\be}{\begin{equation}}
\newcommand{\ee}{\end{equation}\add}
\newcommand{\bea}{\begin{eqnarray}}
\newcommand{\eea}{\end{eqnarray}}

\newcommand{\LCDM}{$\Lambda$CDM\ }

\begin{document}
\begin{center}
{ \bf Averaging in cosmological models} \\[2mm]

\vskip .5in

{\sc A.A. Coley}\\
{\it Department of Mathematics and Statistics}\\
{\it Dalhousie University, Halifax, NS B3H 3J5, Canada}\\
{\it aac@mathstat.dal.ca }
\end{center}

\begin{abstract}

The averaging problem in cosmology is of considerable importance
for the correct interpretation of cosmological data.
We review cosmological observations and discuss some of the issues regarding averaging.
We
present a precise definition of a cosmological model and a rigorous
mathematical definition of averaging, based entirely in terms of scalar invariants.

\end{abstract}



\section{Introduction}

{{Cosmological observations
\cite{Riess}, 
based on
the assumption of a spatially homogeneous and isotropic
Friedmann-Lema\^{i}tre-Robertson-Walker (FLRW) model plus small
perturbations, are usually interpreted as implying that there
exists dark energy, the spatial geometry is flat, and that there
is currently an accelerated expansion, giving rise to the so-called standard
$\Lambda$CDM-concordance model. Although the concordance model is
quite remarkable, it does not convincingly fit all data (see below).
Unfortunately, if the underlying cosmological model
is not a perturbation of an exact flat FLRW solution, the
conventional data analysis and their interpretation is not
necessarily valid.

For example, the standard analysis of type Ia
supernovae (SNIa) and cosmic microwave background (CMB) data in FLRW models cannot be applied
directly when inhomogeneities or backreaction effects are present. 
However, supernovae data can be explained
without dark energy in inhomogeneous models, where the full
effects of general relativity (GR) come into play. 
In one approach exact inhomogeneous cosmological models can be utilised. Indeed, it
has been shown that the Lema\^{i}tre-Tolman-Bondi (LTB)
solution can be used to fit the observed data without the need of
dark energy, although it may be necessary to place the
observer at a preferred location \cite{Celerier}.

A second approach, and the one of interest here, is backreactions through averaging.
The averaging problem in cosmology is of considerable
importance for the correct interpretation of cosmological data.
The correct governing equations on cosmological scales are
obtained by averaging the Einstein field equations (EFE) of GR
(plus a theory of photon propagation; i.e., information on what
trajectories actual particles follow). By assuming spatial
homogeneity and isotropy on the largest scales, the
inhomogeneities affect the dynamics though correction
(backreaction) terms, which can lead to behaviour qualitatively
and quantitatively different from the FLRW models; in particular, the expansion rate may be
significantly affected.}}


\section{Cosmological observations}

From the evidence of the CMB radiation,
the universe was very smooth at the time of last scattering. By the Copernican principle,
the assumption of global isotropy and spatial homogeneity is then justified
at the epoch of last scattering.
Thus, the paradigm for our current standard model of the universe assumes
the underlying geometry is 
FLRW, with additional Newtonian perturbations,
and in
matching the cosmological observables that derive from such a geometry
we have been led to the conclusion over the past decade that the
present--day universe is dominated by a cosmological constant, $\Lambda$, or other
fluid--like ``dark energy'', which
violates the strong energy condition.
In the
case of the \LCDM\ paradigm, dark energy only becomes
appreciable at late epochs.
Dark energy is widely described as the biggest problem in cosmology
today.

There are several problems regarding the  \LCDM model. First, it is difficult to understand 
the large value for $\Lambda$ and
why the contributions of
ordinary matter and the repulsive component are roughly equal
today, at around 10 billion years
(the coincidence problem). 
Second,  the universe is not perfectly homogeneous
and isotropic (or even perturbatively near homogeneity and isotropy).
There are non-linear structures in the real universe which are not described by
perturbations around a smooth background, with a distribution
that is statistically homogeneous and isotropic above a scale of
about 100 Mpc (or, more precisely, 100 $h^{-1}$ Mpc, but we shall omit
the factor $h^{-1}$ for simplicity here) \cite{Hogg}. Linear perturbations
are only valid when both the curvature and the density contrasts remain
small, which is certainly not the case in the non-linear regime of structure formation
when the SNIe observations are made.

Indeed, the largest structures so far detected are limited only
by the size of the surveys that found them
\cite{Labini}.
At the present epoch the distribution of matter is far from
homogeneous on scales less than 150--300 Mpc. The actual universe has a
sponge--like structure, dominated by huge voids surrounded
by bubble walls, and threaded by filaments, within which clusters of galaxies
are located. 
Locally there two enormous voids, both $35$ to $70~\mbox{Mpc}$ across,
associated with the so-called velocity anomaly
\cite{Rizzi:2007th},
a large filament known as the Sloan great wall about
$400~\mbox{Mpc}$ long
\cite{Gott:2003pf} 
and the Shapely
supercluster with a core diameter of $40~\mbox{Mpc}$ at a distance
of $\sim 200~\mbox{Mpc}$
\cite{SSC}. 
In addition, there has been detection of anomolously large local bulk flows
\cite{Kashlinsky} and
evidence for a significant anisotropy
in the local Hubble expansion at distances of $\sim 100~\mbox{Mpc}$
\cite{McClure}. Recent surveys suggest
\cite{HV} 
that some 40--50\% of the present
volume of the universe is in voids of a characteristic scale 30 Mpc.
If smaller minivoids  
and larger supervoids 
are included, then our observed universe is presently void--dominated
by volume; thus within regions as small as $100~\mbox{Mpc}$ density contrasts $\sim -1$
are observed leading to substantial gradients in the (local Ricci) spatial curvature
\cite{Wiltshire}. Therefore, spatial homogeneity is valid only on scales larger than 
at least $100~\mbox{Mpc}$ \cite{Hogg},
in contradiction with the predictions of the  {\LCDM}model in which the scale beyond which the distribution should 
become uniform is about  10 Mpc \cite{Labini}.

The present distribution of matter is clearly very complex, and since
we cannot solve the EFE for this distribution of matter
analytically, there is an important question as to how we operationally
match the average geometry of this distribution to the simple FLRW models.
The mere fact that the universe is presently inhomogeneous means that the
assumptions implicit in the FLRW approximation can no longer be justified
at the present epoch in the almost exact sense that they were justified at
the epoch of last scattering. 
The situation is further complicated by the fact that 
most data analysis based on the standard model (FLRW + perturbations; {\LCDM}) 
is model- and prior- dependent \cite{Shapiro:2005}.

Consequently, spatial homogeneity only applies at the present day
in an averaged sense. 
Given the observed inhomogeneities
and that the nonlinear growth of structure appears to be roughly correlated
to the epoch when cosmic acceleration is inferred to begin,
it has been suggested that the FLRW geometries are inadequate
as a description of the universe at late times and the
introduction of a smooth dark energy is a mistaken interpretation
of the observations. 
A universe which is homogeneous and isotropic only statistically
does not generally expand like an exactly homogeneous and isotropic
universe, even on average. 
It is possible that there are large effects on the 
observed expansion rate (and hence on other observables) due to the
backreaction of inhomogeneities in the universe.  Anything that affects the observed expansion
history of the universe alters the determination of the parameters of dark
energy; in the extreme it may remove the need for dark energy.
Indeed, it has been suggested in 
that inhomogeneities related to structure formation could be
responsible for accelerated expansion \cite{RAS}.

The effects of averaging can be  signicant. Using perturbation theory, effects of order  $\sim 10^{-4}$ are often
quoted. However, these affects occur
by averaging over
the Hubble volumes and not over regions of $\sim 100-200$ Mpc. At best this is  (only) a
self-consistency analysis. In addition, there are highly
non-Gaussian inhomogeneities in the late universe, and the coherence of structures
causes small deviations in observations to sum to a large deviation, and there can be  
significant effects on observations from the backreaction of
inhomogeneities \cite{Kolb}.

In \cite{LiSch}
the hierarchy of the critical
scales for large scale inhomogeneities (backreaction effects) were
calculated, at which $10\%$ effects show up from averaging at different
orders over a local domain in space-time. The dominant contribution comes from the averaged spatial
curvature, observable up to scales of $\sim 200~\mbox{Mpc}$. 
The averaged spatial
curvature typically leads to $10\%$ $(1\%)$ effects
up to $\sim80$ $(240)~\mbox{Mpc}$. The
cosmic variance of the local Hubble rate is $10\%$ $(5\%)$ for
spherical regions of radius 40 $(60)~\mbox{Mpc}$. 
Below $\sim40~\mbox{Mpc}$, the cosmic variance of the Hubble rate is
larger than $10\%$. At lower scales the kinematical backreaction, due to second order perturbations caused by local
inhomogeneities and anisotropies, are important. 
The crude estimates are 
comparable to the actual density variance determined
from large scale structure surveys
\cite{Hogg,Labini,Wiltshire}. In addition,
it has been found that a matter model with discrete masses (rather than an idealised 
continuous fluid) leads to corrections for cosmological parameters
$\sim 10-20 \%$ \cite{Clifton}. Indeed,
it has been argued that the effects of averaging can theoretically 
be as large as $\sim 40 \%$ when the equivalence principle of GR is properly applied \cite{Wiltshire}.

There are also a number of other potential problems with the standard model. 
Apart from WMAP date ($z \sim 1100$),
the standard model is based on local observations ($z < 2$), and consequently
it has been argued that the data does not  convincingly imply acceleration 
 \cite{LiSch}. It is noteworthy that the quality of fit of the {\LCDM}model has
decreased with the introduction of each new SNIa data set, which may hint at inadequacy 
of the {\LCDM} description \cite{RAS}. Indeed,
the standard model does not fit all data;
there is tension  between different SNIa data sets 
\cite{Hicken} and
tension  between different data sets, especially between SNIa data and
 CMB  data \cite{Hicken,Percival},
but also with nucleosynthesis and other large scale structure data
\cite{Wiltshire}.

\subsubsection{Discussion: spatial curvature}

Clearly, backreaction (averaging) effects are real, but their
relative importance still need to be determined. Within perturbation theory, the value of 
the normalized spatial curvature, $\Omega_k$,
is expected to be small. However, 
different authors have argued that $\Omega_k$ can be as large as  $5-10 \%$ 
\cite{Wiltshire, Clifton, RAS}.
In particular, CMB data does not necessarily imply flatness \cite{RAS}; the position of the
CMB peaks is consistent with significant spatial curvature
provided that the expansion history is sufficiently close to the
spatially flat {\LCDM}model.
Indeed, conclusions drawn about spatial curvature from the
CMB are model- and prior-dependent; in a clumpy
universe, the usual expression is inapplicable due to the
non-trivial evolution of the spatial curvature as well as the fact
that clumping contributes to the expansion rate,  and there is no simple argument for
obtaining the position of the CMB peaks. In addition, 
if the spatial curvature (parameter $k$) is
allowed to be a function of position, then considerable spatial
curvature (locally) is permissable (consistent with CMB
observations) \cite{RAS,Clifton}, since curvature can affect different
observations at different scales in different ways (e.g., large scale structure, $z < 2$, and CMB, $z \sim 1100$).

Observational data perhaps suggests a normalized spatial curvature
$|\Omega_{k}| \approx 0.01 - 0.02$ (i.e., of about a percent).
Combining these observations with large scale structure
observations then puts stringent limits on the curvature parameter
in the context of adiabatic $\Lambda$CDM models; however, these
data analyses are very model- and prior-dependent, and care is needed in the proper
interpretation of the data. There is a heuristic argument that
$\Omega_{k}\sim 10^{-3}-10^{-2}$ ($\Omega_k  \sim 1 \%$) \cite{Coley}, which is
consistent with CMB observations
\cite{Riess}  
and agrees with
estimates for intrinsic curvature fluctuations using realistically
modelled clusters and voids in a Swiss-cheese model. 
In particular, the MG equations (see below) were explicitly solved in a FLRW
background geometry and it was found that the correlation tensor
(backreaction) is of the form of a spatial curvature \cite{CPZ}. Thus, the
averaged EFE for a flat spatially homogeneous,
isotropic macroscopic space-time geometry has the form of the
EFE of GR for a non-flat spatially homogeneous,
isotropic space-time geometry.

It must be
appreciated that such a value for $\Omega_{k}$, at the 1\% level,
is relatively large and may have a significant dynamical effect on
the evolution of the universe and the interpretation of
cosmological observations.
Indeed, in such a scenario the
current contribution from the spatial curvature is  comparable to the
energy density in luminous matter. In addition, such a value
cannot be naturally explained by inflation. From standard
analysis, depending on the initial conditions and the details of a
specific model of inflation, $|\Omega - 1|$ would be extremely
small. 
Therefore, any value for $\Omega_{k}$ at the 1 \% level  can only
be naturally explained in terms of an averaging effect.  In addition, such an
effect would compromise any efforts to use data to constrain dark energy models
(within the standard paradigm) with a variable equation of state \cite{SSS2}.

\section{The averaging problem in cosmology}

\subsection{General Approaches}

The Universe is not isotropic
or spatially homogeneous on local scales. 
The gravitational FE on large scales are obtained by
averaging the EFE of GR. It is
necessary to use an exact covariant approach which gives a
prescription for the correlation functions that emerge in an
averaging of the full tensorial EFE. 

There are a number of approaches to the averaging problem. In the approach of Buchert a
3+1 cosmological space-time splitting is employed (i.e., this procedure is not generally 
covariant) and
only scalar
quantities are averaged (and thus the governing equations are not closed) \cite{Buch}. The perturbative approach 
(backreaction about an FLRW background  \cite{Kolb}) involves
averaging the perturbed EFE. However, a
perturbation analysis cannot provide any information about an
averaged geometry; thus perturbation
theory cannot be conclusive and provide a complete solution.

To date the macroscopic gravity (MG)  approach is the
only approach to the averaging problem in GR which
gives a prescription for the correlation functions which emerge in
an averaging of the non-linear FE (without which the
averaging of the EFE simply amount to definitions
of the new averaged terms) \cite{Zal}. The MG space-time averaging approach is a fully covariant,
gauge independent and exact method, in which the averaged EFE
are written in the form of the EFE
for the macroscopic metric tensor when the correlation
terms are moved to the right-hand side of the averaged EFE
to serve as the geometric modification to the averaged
(macroscopic) matter energy-momentum tensor. For the cosmological
problem additional assumptions are required: with reasonable
cosmological assumptions, the correlation tensor in Zalaletdinov's scheme
takes the form of a spatial curvature
\cite{CPZ}, 
and Buchert's scheme
can be realized as a consistent limit \cite{PS}.

There are other approaches to averaging.
The formal mathematical issues of averaging tensors on a
differential manifold has recently been revisted.
We note that integrating scalars on spacetime regions is
always well-defined and it may be possible to
avoid several of the technical problems of averaging by adopting an
approach based  on scalar curvature invariants.

\section{Cosmological models}

A cosmological model
is a {\em mixed} model, in that the matter is already assumed to
be averaged but the geometry is not (necessarily). Therefore, we need a
consistent model for the matter, represented on the characteristic
averaging scale, and its appropriate (averaged) physical
properties. It is known that the separation between the
gravitational field and the matter is not scale invariant and the
notion of a perfect fluid is not scale invariant;
averaging (in the presence of a gravitational field) modifies the
equation of state of the matter. In addition, since
averaging does not conserve geodesics, we need further assumptions
in order to be able to compare the models with observational data.

A precise definition of a cosmological model is necessary;
i.e., a framework in which to do averaging.
The definition we shall adopt is given by the
following conditions C1 -- C5 \cite{Scalar}:
{\bf{C1. Spacetime Geometry:}} The spacetime geometry
$({\bf M,g})$  is defined by a smooth Lorentzian  metric ${\bf g}$
(characterizing the macroscopic gravitational field) defined on a
smooth differentiable manifold ${\bf M}$.
The macroscopic metric  geometry is obtained by an appropriate
spacetime averaging of the microgeometry; thus part of the
definition  of a cosmological model consists of specifying {\em {the
averaging scheme}} (which must be consistent with the physical
assumptions of the model encapsulated in the conditions C3 and C4
below) and {\em {the cosmological scale}} over which averaging or the
smoothing occurs (i.e., we must specify the averaging scale $\ell$
or averaging region).

{\bf{C2. Timelike Congruence:}} There exists a timelike
congruence $({\bf u})$ (in principle locally), representing a family of
fundamental observers.
Mathematically this means that the spacetime is
\emph{$\mathcal{I}$-non-degenerate} and hence the spacetime is
uniquely characterized by its scalar curvature invariants \cite{INV}.
In addition to the formal parts C1 and C2 of the definition of a
cosmological model $({\bf M,g, \ell, u})$, we must also specify
the physical relationship (interaction) between the macroscopic
geometry and the matter fields, including how the matter responds
to the macroscopic geometry.

{\bf{C3. Macroscopic FE:}} There exists an
appropriate set of macroscopic FE relating the
averaged matter and appropriately averaged (or macroscopic)
geometry. This is based on an underlying microscopic theory of
gravity (such as, for example, GR), and an appropriate formalism
to average the geometry and find corrections (correlations) due to
averaging the Einstein tensor in the resulting FE:

\begin{equation}
\tilde{G}^{a}_{~b} + C^{a}_{~b}=T^{a}_{~b},
\end{equation}
where $\tilde{G}^{a}_{~b} \equiv \tilde{R}^{a}_{~b} -
\frac{1}{2}{\delta}^{a}_{~b}\tilde{R}$ and $\tilde{R}^{a}_{~b}$ is
the Ricci tensor of the averaged macrogeometry, $C^{a}_{~b}$ is
the correlation tensor, and $T^{a}_{~b}$ is the energy momentum
tensor (already assumed averaged).

{\bf{C4. Equations of motion:}} We also need to know the
trajectories along which the cosmological matter moves (and also
the light trajectories, which determine observational relations).
In principle, the
average motion of a photon need not be a null geodesic in the
averaged geometry \cite{Coley}.
{\bf{C5. Observations:}} Finally, we need to be able to
relate averaged quantities with physical observables, which
ultimately must be consistent with cosmological data.

In the standard FLRW model there are a number of simplifications
and assumptions.
The past approaches to averaging have been ideally suited to the
FLRW models (with small, vanishing in the limit, perturbations).
In these models, the macrometric ${\bf g}$ is the FLRW metric
(C1) and ${\bf u}$ also has a geometric meaning (C2). In the usual
point of view there are no correlations due to averaging (i.e.,
$C^{a}_{~b}=0$) or, more
precisely, they are negligible (C3). In this case it follows from
the contracted Bianchi identities that energy-momentum is
conserved: $T^{c}_{~b;c}=0$, which relates the matter to the
averaged geometry. All other effects are assumed negligible (C4).
However, there is no formal argument that such assumptions arise from
a rigorous averaging scheme of some appropriate (physically
motivated) microgeometry. In addition, there are some important
effects in the standard model which are not necessarily small perturbations.

Since there are no scales explicitly specified in the model, in a sense the
model is incomplete. Indeed, the model does not even have the
ability to determine whether there is a scale above which the
geometry is exactly FLRW or whether at all scales the geometry is
only approximately FLRW (with a given perturbation scale). Furthermore, 
regarding C5, we can ask whether the model agrees with
observations? If it does not, then even if the model agrees in
some approximate sense with most observations, there is no
structure within which to discuss the potential small
discrepancies with observed data, which is a deficiency of the
model. If the model does, then it would be remarkable, although
there is still the need for a physical explanation for the dark
energy. Finally, if observations indicate that $\Omega_k \approx 1-2\%$, 
then there is no physical mechanism
within the model (particularly if there is an inflationary period)
to produce an intrinsic curvature parameter $k$ of this magnitude,
whereas an effective curvature parameter $\hat{k}$ of about a
percent arises naturally from averaging.

\section{An approach to averaging using scalars}

For any given spacetime $({\bf M,g})$ we define the set of all
scalar polynomial curvature invariants
\begin{equation} \label{eqn}
\mathcal{I} \equiv \{R,R1,R2,R3, R^2,R^{\mu}_{~\nu}R^{\nu}_{~\mu},\dots, C^2, \dots\}
\end{equation}
(where the $Ri$ are eigenvalues of the Ricci tensor, and  $C^2 \equiv
C^{\mu\nu\alpha\beta} C_{\mu\nu\alpha\beta}$).
Consider a spacetime $({\bf M,g})$ with a set of
invariants $\mathcal{I}$. Then, if there does not exist a
continuous metric deformation of $g$ having the same set of
invariants as $g$, the set of invariants will be called 
\emph{non-degenerate}, and the spacetime metric, $g$, 
\emph{$\mathcal{I}$-non-degenerate}. This implies that
for a metric which is $\mathcal{I}$-non-degenerate the invariants
characterize the spacetime uniquely, at least locally. 
It was proven  \cite{INV}
that a 4D spacetime is either $\mathcal{I}$-non-degenerate or the metric is a 
degenerate Kundt metric. This is a
striking result because it tells us that the only metrics not locally determined by
their scalar invariants must be of Kundt form.

Hence, in general, since we know how to average scalar
quantities, we can average all of the scalar curvature invariants
that then represents an averaged spacetime (with that set of
averaged scalar invariants). 
In particular, we note that cosmological models (as defined above) belong to the set of
spacetimes completely characterized by their scalar 
invariants, suggesting that we can average a cosmological model
using scalar invariants. Therefore, we have a microgeometry completely
characterized by its set of scalar curvature invariants
$\mathcal{I}$. We then average these microgeometry scalar
curvature invariants to obtain a new set of macrogeometry scalar
curvature invariants  $\tilde{\mathcal{I}}$, which now completely
characterizes the macrogeometry \cite{Scalar}.

\subsection{Averaging the geometry}

In the general mathematical context we want to describe the
averaged geometry (represented by the Riemann tensor and its
covariant derivatives) and interpret the results. Let us consider,
$\mathcal{I}$, defined by (\ref{eqn}), which
is an ordered set of functions on {\bf M}. Let us we write
$\tilde{\mathcal{I}}\equiv\{\tilde{R},\dots,\widetilde{R^{\mu}_{~\nu}R^{\nu}_{~\mu}},\dots\}$,
which is also an ordered set of functions.
The question is then: does the ordered set of
functions $\tilde{\mathcal{I}}$ correspond to the associated
scalar curvature invariants for some metric $\tilde{g}$ (which
could then serve to {\em define} the macrometric $\tilde{g}$).

It is certainly plausible that (some
appropriately defined subset of) the ordered set of functions
$\tilde{\mathcal{I}}$ correspond to the associated scalar
curvature invariants for some macrometric $\tilde{g}$ for the
class of $\mathcal{I}$-non-degenerate geometries that constitute
the class of cosmological models defined. Since the
geometries are $\mathcal{I}$-non-degenerate and in 4D the properties
of the geometry can be represented in terms of scalars, and since
relations between different terms (functions) in the set
$\mathcal{I}$ (e.g., $R$ and $R^2$ are functionally dependent) and
the corresponding terms in the set $\tilde{\mathcal{I}}$ (e.g.,
$\tilde{R}$ and $\tilde{R}^2$) are functionally related in exactly
the same way and syzygies (e.g., describing the algebraic type)
are maintained under averaging, it follows that in general the set
$\tilde{\mathcal{I}}$ gives rise to a macrometric
$\tilde{g}$ (which will have similar algebraic properties to the
micrometric ${g}$).

\subsubsection{Proposal: Scalar Averaging Procedure}

Let us consider the ordered set of functions
$\mathcal{I}$ in the form of (\ref{eqn}).
First, let us omit any scalars from this set that are not algebraically independent
(e.g., $\{R^2,R^{\mu}_{~\nu}R^{\nu}_{~\mu}\dots\}$)
 to obtain an
(appropriate `independent') subset $\mathcal{I}_A$. Second, 
for a particular spacetime, we omit any scalars from $\mathcal{I}_A$
that can be obtained from syzygies defining that particular spacetime
(e.g., defining the algebraic type of the spacetime, such as the Segre type or 
the Petrov type). For example, for a Ricci tensor corresponding to the algebraic
form of a perfect fluid we could omit $\{R2,R3\}$ (relative to $\{R,R1\}$). We consequently obtain the
subset $\mathcal{I}_{SA}$: e.g., 
$\mathcal{I}_{SA} \equiv \{R,R1,\dots, C^2,\dots\}.$
For the spacetimes under consideration the
microgeometry is then completely characterized by the (sub)set of scalar
curvature invariants $\mathcal{I}_{SA}$.

We now construct the new ordered set of functions $\tilde{\mathcal{I}}_{SA}$ by averaging
the various scalar invariants of $\mathcal{I}_{SA}$:
$\tilde{\mathcal{I}}_{SA}\equiv\{\tilde{R}, \widetilde{R1}, \dots, \widetilde{C^2},\dots\}$,
where all of the original scalar invariants omitted from the original set $\mathcal{I}$
are replaced by a new set of functions obeying exactly the same algebraic properties
(or syzygies)  as $\mathcal{I}_{SA}$. Therefore, it is assumed that $\tilde{\mathcal{I}}_{SA}$
comes equiped with these syzygies, so that we could construct the corresponding set $\tilde{\mathcal{I}}$
consisting of the members of $\tilde{\mathcal{I}}_{SA}$ and all of the corresponding syzygies.
Consequently, the set $\mathcal{I}_{SA}$
is an ordered set of functions (scalar curvature invariants) on {\bf M} which 
uniquely determines the macrogeometry
with exactly the same algebraic properties as the original microgeometry.

\subsubsection{Cosmological models}

In the case of a cosmological model, from C3 we have an effective set of FE
and we only need to consider the macrogeometric Ricci tensor
$\tilde{R}^{a}_{~b}$ (the correlation tensor is obtained from
the averaging procedure). The microgeometric Ricci tensor
${R}^{a}_{~b}$ is completely characterized by a set of scalar
curvature invariants $\mathcal{I}_R$. Averaging these scalar
curvature invariants we obtain the set
$\tilde{\mathcal{I}}_{\tilde{R}}$, which completely characterizes
the macrogeometric Ricci tensor $\tilde{R}^{a}_{~b}$. Since
constructing the Ricci tensor from a set of scalar curvature
invariants $\mathcal{I}_R$ is relatively simple compared to the
corresponding problem for the Riemann tensor, and since the
reduced set of scalar curvature invariants $\mathcal{I}_R$ is
considerably smaller than $\mathcal{I}$, we have 
reduced the complexity of the problem in this new averaging
approach. Indeed, for a Ricci tensor of the algebraic form of a
perfect fluid, there are effectively (only) two independent zeroth order scalar
invariants, the Ricci scalar and a single Ricci eigenvalue
(corresponding to the effective energy density, $\rho$, and
pressure, $p$, of the perfect fluid). Therefore, in the context
of the scalar averaging procedure, we have the set
$\{\tilde{R}, \widetilde{R1}\}$.

It is necessary to determine whether the correlations due
to averaging alter the geometry or affect the effective
energy-momentum tensor. This is partly a question of
interpretation, which must be done within the context of the
underlying cosmological model.
In particular, in the cosmological application it may be
appropriate to reinterpret the averaging correlations as
corrections to the matter fields (and hence the effective equation of state) 
through the EFE.

In \cite{INV} the specific example of a static spherically
symmetric perfect fluid spacetime
was considered. This is a simple and
appropriate model for illustration since it can include an
arbitrary function of one variable, there is a non-vanishing
pressure, the averaging region does not change with time and there
are no gravitational waves. The average correlations can be interpreted as contributing a
small constant curvature term,
arising from the averaging of local inhomogeneities in the
micro-Ricci tensor to the smooth macro-Ricci tensor (consistent
with the results of \cite{CPZ}).

{\em Acknowledgements}. 
This work was supported by NSERC of Canada.

\end{document}